\documentstyle[12pt,amssymb]{article}

\newtheorem{lemma}{Lemma}
\newtheorem{proposition}{Proposition}

\begin{document}

\begin{center}

{\Large \bf
Equilibrium of charges and differential equations solved by polynomials
}\\*[3ex]
{\large \bf Igor Loutsenko}\\*[1ex]
SISSA, Via Beirut 2-4, 34014, Trieste, Italy \\
e-mail: loutseni@fm.sissa.it
\end{center}

\begin{center} \bf Abstract \end{center}

\begin{quote}

We examine connections between rationality of certain indefinite integrals and
equilibrium of Coulomb charges in the complex plane .

\end{quote}

\section*{I. Introduction. Adler-Moser Polynomials}

In their 1929 paper  J.L.Burchnall and T.W. Chaundy \cite{BC} examined
the following (apparently elementary) question: 
What condition must be satisfied by two polynomials $p(z)$ and $q(z)$ 
in one variable $z$ in order the indefinite integrals
\begin{equation}
\int\left(\frac{p(z)}{q(z)}\right)^2dz, \quad \int\left(\frac{q(z)}{p(z)}\right)^2dz
\label{rational}
\end{equation}
may be rational, i.e. expressible without logarithms. Under the assumption 
that $p$ and $q$ have no common or repeated factors the problem is reduced 
to finding polynomial solutions of the following differential-recurrence equation
\begin{equation}
p^{\prime\prime}q-2p^\prime q^\prime+pq^{\prime\prime}=0
\label{minusplus}
\end{equation}
It was shown that solutions to (\ref{minusplus}) are Adler-Moser polynomials \cite{KDV}:
\begin{equation}
p=\theta_i,\quad q=\theta_{i+1}, \quad \theta_0(z)=1,\quad \theta_1(z)=z,
\quad \deg \theta_i =\frac{i(i+1)}{2}
\label{adler}
\end{equation}
The recurrence relation (\ref{minusplus}) for 
the Adler-Moser polynomials (\ref{polynomials}) can be rewritten as the first-order
(three term) relation \cite{KDV}, \cite{BC}
\begin{equation}
\theta^\prime_{n+1}\theta_{n-1}-\theta_{n+1}\theta^\prime_{n-1}=(2n+1)\theta^2_n
\label{three}
\end{equation}
Considering it as a first order linear differential equation for $\theta_{n+1}$ we
obtain recursively all solutions introducing an integration constant at each step.
Thus the $n$th polynomial depends on $n$ free parameters
$$
\theta_n=\theta_n( z; t_1, t_2,\dots , t_n)
$$
Polynomiality of all solutions is assured by the rationality of (\ref{rational}).

Several first examples of the Adler-Moser polynomials are:
\begin{equation}
\begin{array}{l}
\theta_0=1 \\
\theta_1=z \\
\theta_2=z^3+t_2 \\
\theta_3=z^6+5t_2z^3+t_3z-5t_2^2 \\
...
\end{array}
\label{kdv}
\end{equation}
In (\ref{kdv}), we set $t_1=0$ , since $t_1$ can be absorbed by the translation $z \to z+t_1$

It was observed in \cite{Bartman} that the roots of the consecutive polynomials $\theta_i, \theta_{i+1}$
 are the equilibrium coordinates of $i(i+1)/2$ positive and 
$(i+1)(i+2)/2$ negative Coulomb charges (with values $\pm 1$ correspondingly) on the
plane. Namely, the function
$$
E=\sum_{i<j=1}^{n(n+1)/2}\ln|x_i-x_j|+\sum_{i<j=1}^{(n+1)(n+2)/2}\ln|y_i-y_j|-
\sum_{i=1}^{n(n+1)/2}\sum_{j=1}^{(n+1)(n+2)/2}\ln|x_i-y_j|
$$ 
has a critical point when $x_i, y_i$ are roots of $\theta_n, \theta_{n+1}$ respectively.

The constructive way to find the Adler-Moser polynomials is to use their determinant
representation built by means of the Darboux-Crum transformations of the operator
$d^2/dz^2$.

This construction can be demonstrated in the following diagram:
\begin{equation}
\to H_i=L_iM_i=M_{i+1}L_{i+1} \to H_{i+1}=L_{i+1}M_{i+1}=M_{i+2}L_{i+2} \to H_{i+2} ...
\label{Darboux}
\end{equation}
where
$$
L_i=\frac{\theta_i(z)}{\theta_{i-1}(z)}\frac{d}{dz}\frac{\theta_{i-1}(z)}{\theta_i(z)},
\quad
M_i=\frac{\theta_{i-1}(z)}{\theta_i(z)}\frac{d}{dz}\frac{\theta_{i}(z)}{\theta_{i-1}(z)}
$$
and $L_0=M_0=d/dz$.

From (\ref{Darboux}), we derive the intertwining differential operators $D_i, U_i$
\begin{equation}
U_i\frac{d^2}{dz^2}=H_iU_i, \quad \frac{d^2}{dz^2}D_i=D_iH_i \qquad D_i=M_i ... M_1,
\quad U_i=L_1 ... L_i
\label{up}
\end{equation}
connecting $H_0=d^2/dz^2$ with $H_i$.

This, together with the obvious property
$
D_iU_i=\frac{d^{2i}}{dz^{2i}}
$
leads to the Wronskian representation for the Adler-Moser polynomials. In more details
\begin{equation}
\theta_n={\rm const}W[\psi_1,\dots,\psi_n],\quad \psi_n^{\prime\prime}=\psi_{n-1},
\quad \psi_1=z
\label{polynomials}
\end{equation}
where the $n$th polynomial is (up to a constant factor) the
Wronskian $W=\det||d^i\psi_j/dz^i||_{0\le i \le n-1, 1\le j\le n}$.

The intertwining operators are also expressible in terms of Wronskians. For instance
\begin{equation}
U_i[\psi]=W[\psi_1,\dots,\psi_i,\psi]/W[\psi_1,\dots,\psi_i]
\label{shift}
\end{equation}

\section*{II. Indefinite integrals related to equilibrium of charges}

By analogy with the previous section we pose the following question:
when does an equilibrium condition for charges of different sign is equivalent
 to rationality of some indefinite integrals?

Before addressing this question we write the equilibrium equation for two types of
charges in terms of polynomials:
\begin{lemma}
Let
\begin{equation}
E=\sum_{1\le i<j\le n+m} Q_iQ_j\ln|z_i-z_j|
\label{energy}
\end{equation}
be a real function of $n+m$ complex
variables $z_i \in {\Bbb C}, i=1 \dots n+m$ and
$$
Q_i \in {\Bbb R}, \qquad Q_i=\left\{\begin{array}{l} 1, \qquad i=1 \dots n \\
-\Lambda, \qquad i=n+1 \dots m+n \end{array}\right.
$$
Then $E$ has a critical point at
$$
z_i=\left\{\begin{array}{l} x_i, \qquad i=1 \dots n \\ y_{i-n}, \qquad i=n+1
\dots m+n \end{array}\right.
$$
iff
\begin{equation}
\left\{p,q\right\}_\Lambda:=\frac{d^2p(z)}{dz^2}q(z)-
2\Lambda\frac{dp(z)}{dz}\frac{dq(z)}{dz}+\Lambda^2p(z)\frac{d^2q(z)}{dz^2}=0
\label{second}
\end{equation}
where $p, q$ are the following polynomials
$$
p(z)=\prod_{i=1}^n(z-x_i), \quad q(z)=\prod_{i=1}^m(z-y_i)
$$
\label{lemmaeq}
\end{lemma}

Proof: Is a calculation. We use the partial-fraction decomposition of $\frac{\{p,q\}_\Lambda}{pq}$ around
singular points $x_i$ and $y_i$.

Now we are in position to find an equivalent condition for $p$ and $q$ in terms of
rational integrals. 

Indefinite integrals of the algebraic function do not contain logarithmic terms if the
function does not have singularities of the type $z^{-1}$. The following lemma shows
when certain functions, related to solutions of (\ref{second}) ,  are free of simple poles and their integrals are rational.

\begin{lemma}
Let $ p(z)=\prod_{i=1}^n(z-x_i), \quad q(z)=\prod_{i=1}^m(z-y_i) $ do not have multiple
or common roots. Then residues of simple poles of $q^{2\Lambda}/p^2$ and $p^{2/\Lambda}/q^2$ vanish:
$$
Res_{z=x_i}\frac{q(z)^{2\Lambda}}{p(z)^2}=0, \quad i=1\dots n, 
\quad Res_{z=y_i}\frac{p(z)^{2/\Lambda}}{q(z)^2}=0, \quad i=1\dots m
$$
iff $p$ and $q$ satisfy (\ref{second}).

Both indefinite integrals
\begin{equation}
\int\frac{q(z)^{2\Lambda}}{p(z)^2}dz, \quad \int\frac{p(z)^{2/\Lambda}}{q(z)^2}dz
\label{intesecond}
\end{equation}
are rational iff $p,q$ satisfy (\ref{second}) with $\Lambda=1/2, 1, 2$
\label{lemmaint}
\end{lemma}

Proof: Let us factor $p(z)$ as $p(z)=(z-x_i)P(z)$. Then the condition $Res_{z=x_i}\frac{q(z)^{2\Lambda}}{p(z)^2}=0$
implies that
$$
\frac{d}{dz}\left(\frac{q(z)^{2\Lambda}}{P(z)^2}\right)_{z=x_i}=0
$$
Since $P(x_i)=p^\prime(x_i)$, $P^\prime(x_i)=p^{\prime\prime}(x_i)/2$ we get the following equation
$$
p^{\prime\prime}(x_i)q(x_i)-2\Lambda p^\prime(x_i)q^\prime(x_i)=0
$$
Thus, since $x_i$ are roots of $p(z)$
\begin{equation}
p^{\prime\prime}(z)q(z)-2\Lambda p^\prime(z)q^\prime(z)=a(z)p(z)
\label{pq}
\end{equation}
where $a(z)$ is a polynomial.

Similarly, we get from the condition $Res_{z=y_i}\frac{p(z)^{2/\Lambda}}{q(z)^2}=0$ that
\begin{equation}
\Lambda^2 p(z)q^{\prime\prime}(z)-2\Lambda p^\prime(z)q^\prime(z)=b(z)q(z)
\label{qp}
\end{equation}
Adding $ \Lambda^2 p(z)q^{\prime\prime}(z) $ to (\ref{pq}) and $ p^{\prime\prime}q(z) $ to (\ref{qp}) we get
$$
p^{\prime\prime}(z)q(z)-2\Lambda p^\prime(z)q^\prime(z)+\Lambda^2p(z)q^{\prime\prime}(z)=A(z)p(z)=B(z)q(z)
$$
where $A(z)=a(z)+\Lambda^2 q^{\prime\prime}(z), B(z)=b(z)+p^{\prime\prime}(z)$.

Since $p(z)$ and $q(z)$ do not have common roots it follows from
$$
A(z)p(z)=B(z)q(z)
$$
that $ A(z)=N(z)q(z), B(z)=N(z)p(z)$, where $N(z)$ is a polynomial. Thus, we arrive at
\begin{equation}
p^{\prime\prime}(z)q(z)-2\Lambda p^\prime(z)q^\prime(z)+\Lambda^2p(z)q^{\prime\prime}(z)=N(z)p(z)q(z)
\label{zerorhs}
\end{equation}
The degree of the polynomial in the left hand side of (\ref{zerorhs}) is at most $\deg(q)+\deg(p)-2$ while
the degree of the right hand side is at least $\deg(q)+\deg(p)$, unless $N(z)=0$. Therefore, $N(z)=0$ and we
obtain (\ref{second}). The choice $\Lambda=1/2, \Lambda=1, \Lambda=2$ is essential for the rationality of (\ref{intesecond})
because the both exponents $2\Lambda$ and $2/\Lambda$ are integer. This completes the proof. 

Since the cases $\Lambda=1/2$ and $\Lambda=2$ are equivalent (one is connected with the other
by permutation of $p$ with $q$) we have two second order bilinear differential relations
corresponding to two types of rational integrals

\begin{itemize}

\item Adler-Moser polynomials, $\Lambda=1$ (see \cite{BC})

\begin{equation}
\begin{array}{l}
\{p,q\}_1=\frac{d^2p(z)}{dz^2}q(z)-2\frac{dp(z)}{dz}\frac{dq(z)}{dz}+p(z)
\frac{d^2q(z)}{dz^2}=0\\
\\
\int\frac{q(z)^2}{p(z)^2}dz, \quad \int\frac{p(z)^2}{q(z)^2}dz
\end{array}
\label{adler_moser}
\end{equation}

\item $\Lambda=2$,

\begin{equation}
\begin{array}{l}
\{p,q\}_2=\frac{d^2p(z)}{dz^2}q(z)-4\frac{dp(z)}{dz}\frac{dq(z)}{dz}+
4p(z)\frac{d^2q(z)}{dz^2}=0\\
\\
\int\frac{q(z)^4}{p(z)^2}dz, \quad \int\frac{p(z)}{q(z)^2}dz
\end{array}
\label{bartman}
\end{equation}

\end{itemize}

Equation $\{p,q\}_2=0$ has been mentioned in \cite{Bartman} in connection with the problem of equilibrium of the point vortices in two-dimensional hydrodynamics.

We devote the sequel of the work to investigation of the second case and its
generalizations.

\section*{III. $ \Lambda=2 $ }

We begin by summing up what we have obtained so far.

Let $p(z)$ and $q(z)$ be polynomials of the complex variable $z$ which do not have
common or multiple roots.

We study the following three problems:

\begin{itemize}

\item When the indefinite integrals
\begin{equation}
\int\frac {p} {q^2}dz, \quad \int\frac{q^4}{p^2}dz
\label{integrals1}
\end{equation}
are rational ?

\item When the system of $n$ and $m$ Coulomb charges of values $1$ and $ -2$
respectively
can be in equilibrium in the complex plane and what are coordinates of the charges ?

In other words when does the energy function
\begin{equation}
E=\sum_{i<j}Q_iQ_j\log|z_i-z_j|, \quad Q_i=\left\{\begin{array}{l} 1,\quad i=1..n\\
-2,\quad i=n+1..n+m \end{array}\right.
\label{energy1}
\end{equation}
have a critical point and at which $z_i,i=1..n+m$ ?

\item When does the equation
\begin{equation}
\left\{p,q\right\}_2=0 
\label{recurrence1}
\end{equation}
have polynomial solutions?

\end{itemize}

The answer to questions (\ref{integrals1}), (\ref{energy1}) and (\ref{recurrence1}) is given by

\begin{proposition}
Let $i \in {\Bbb Z}$, and (modulo translation of $z$) $p_{-1}=z,\quad p_0=q_0=1,
\quad q_1=z$. 
Then all solutions of the recurrence relations
\begin{equation}
\begin{array}{l}
q_{i+1}^\prime q_i-q_{i+1}q_i^\prime=(3i+1)p_i\\
p_i^\prime p_{i-1}-p_ip_{i-1}^\prime=(6i-1)q_i^4
\end{array}
\label{free}
\end{equation}
are polynomials of the degrees
$$
\deg(p_i)=i(3i+2),\quad \deg(q_i)=i(3i-1)/2, \quad i \in {\Bbb Z}
$$
They provide all solutions of (\ref{recurrence1})
$$
\left\{p_i,q_i\right\}_2=\left\{p_i,q_{i+1}\right\}_2=0, \quad i \in {\Bbb Z}
$$
Integrals (\ref{integrals1}) are rational iff $p$ and $q$ are connected by the bilinear equation
(\ref{recurrence1}) i.e. iff
$$
p=p_i, q=q_i, \quad {\rm or} \quad p=p_i, q=q_{i+1}
$$
The energy function (\ref{energy1}) has a critical point provided $z_i$ are zeros of
above pairs of polynomials: zeros of $p$, and $q$ being positions of charges 1 and -2 respectively.
\label{nofield}
\end{proposition}

Proof:

\begin{itemize}

\item The equivalence between the bilinear equation (\ref{recurrence1})
 and rationality of integrals (\ref{integrals1})
is a corollary of Lemma \ref{lemmaint}.

\item The equivalence between  (\ref{recurrence1})
and existence of critical points of the energy function (\ref{energy1})
is a corollary of Lemma \ref{lemmaeq}.

\item Since $p=z^n+\dots, \quad q=z^m+\dots$, from the highest symbol of
(\ref{recurrence1}) we get the Diophantine equation connecting $n$ with $m$
$$
(n-2m)^2-n+4m=0
$$
Its solutions are
$$
n=n_i, m=m_i, \quad {\rm or} \quad n=n_i, m=m_{i+1}, \quad i \in {\Bbb Z}
$$
where
$$
n_i=i(3i+2), \quad m_i=\frac{i(3i-1)}{2}
$$
Let $\{p_i,q_i\}_2=0$. Considering this equation as a second order differential equation
with a solution $q_i$, by elementary methods we find that
its second linearly independent solution   is given by
\begin{equation}
q_{i+1}=(3i+1)q_i\int\frac{p_i}{q_i^2}dz, \quad \{p_i,q_{i+1}\}_2=0
\label{qup}
\end{equation}
and is polynomial by the rationality of (\ref{integrals1}). The degrees of polynomials are
connected by
$$
\deg(q_{i+1})+\deg(q_i)=\deg(p_i)+1
$$ 
It is seen that this relation is satisfied if $\deg(p_i)=n_i, \deg(q_i)=m_i$.

The similar procedure holds if we fix $q_i$ and consider $p_{i-1}$ and $p_i$ as linearly
independent solutions of (\ref{recurrence1}):
\begin{equation}
p_i=(6i-1)p_{i-1}\int\frac{q_i^4}{p_{i-1}^2}dz
\label{pup}
\end{equation}

By freedom in choosing linearly independent solutions of (\ref{recurrence1})
we write analogs of (\ref{qup}) and (\ref{pup}) for decreasing indices
\begin{equation}
q_i=-(3i+1)q_{i+1}\int\frac{p_i}{q_{i+1}^2}dz, \quad
p_{i-1}=-(6i-1)p_i\int\frac{q_i^4}{p_i^2}dz
\label{opposite}
\end{equation}
and we can generate $p_i, q_i$ by induction in either directions starting at some $i$.

Since $n_i$ and $m_i$ are strictly increasing for $i\ge 0$ (strictly decreasing for
$i\le 0$) and $m_0=n_0=0$, the induction terminates for these two branches at
$p_0=q_0=1$.

Rewriting  (\ref{qup}), (\ref{pup}) or (\ref{opposite}) in the differential form, we get
the first-order recurrence relations (\ref{free}), which completes the proof.

\end{itemize}

Here we write down several examples of polynomials satisfying above conditions for the
branch $i\ge 0$
$$
\begin{array}{ll}
q_0=1 & \quad p_0=1\\
q_1=z & \quad p_1=z^5+t_1\\
q_2=z^5+\tau_2z-4t_1 & \quad p_2=z^{16}+... \\
\dots & \quad \dots
\end{array}
$$
and for the branch $i\le 0$
$$
\begin{array}{lll}
p_0=1 & \quad & q_0=1 \\
p_{-1}=z & \quad & q_{-1}=z^2+\tau_{-1} \\
\begin{array}{l}
p_{-2}=z^8+\frac{28}{5}\tau_{-1}z^6+14\tau_{-1}^2z^4\\+28\tau_{-1}^3z^2+t_{-2}z-7\tau_{-1}^4
\end{array}
& \quad &
\begin{array}{l}
\quad q_{-2}=z^7+7\tau_{-1}z^5+35\tau_{-1}^2z^3+\tau_{-2}z^2-35\tau_{-1}^3z\\+\tau_{-1}\tau_{-2}-\frac{5}{2}t_{-2} \\
\end{array} \\
\dots & \quad \dots
\end{array}
$$
where $t_i$ and $\tau_i$ stand for arbitrary constants.

\section*{IV. Intertwining and factorization}

One can try to find polynomials for $\Lambda=2$ explicitly by analogy with the
Adler-Moser case.

It easy to observe that equation (\ref{recurrence1}) can be written in two different
Scr\"odinger forms
\begin{equation}
\begin{array}{ll}
S\left[\frac{p}{q^2}\right]=\left(\frac{d^2}{dz^2}+6(\ln q)^{\prime\prime}\right)
\left[\frac{p}{q^2}\right]=0 & \qquad {\rm (a)} \\
\tilde S\left[\frac{q}{\sqrt{p}}\right]=\left(\frac{d^2}{dz^2}
+\frac{3}{4}(\ln p)^{\prime\prime}\right)\left[\frac{q}{\sqrt{p}}\right]=0 &
\qquad {\rm (b)}
\end{array}
\label{s}
\end{equation}
connected by the following factorization
$$
\frac{p}{q^2}S=LM, \quad \frac{p}{q^2}\tilde S=ML,
\quad L=\frac{d}{dz}\frac{\sqrt{p}}{q},
\quad M=\frac{p^{3/2}}{q^3}\frac{d}{dz}\frac{q^2}{p}
$$
Since $p=p_{i-1}$ and $p=p_{i}$ are solutions of (\ref{recurrence1}) with $q=q_i$,
we can put any of them in (\ref{s} a) with $q=q_i$.
The similar fact holds for (\ref{s} b), with $q=q_i$ or $q=q_{i+1}$ and $p=p_i$.
This ambiguity in the choice of solutions
results in a possibility of different factorizations for the same second order operator.
We remind that a similar possibility led to explicit representation of the Adler-Moser polynomials for $\Lambda=1$.
 
Now, the factorization diagram has the following form
\begin{equation}
\begin{array}{cccccc}
\downarrow & & & & & \\
\frac{p_{i-1}}{q_i^2}S_i = \tilde L_i \tilde M_i& \rightarrow &
\frac{p_i}{q_i^2}S_i = L_i M_i & & & \\
& & \downarrow & & &  \\
& & \frac{p_i}{q_i^2}\tilde S_i =  M_i L_i & \rightarrow &
\frac{p_i}{q_{i+1}^2}\tilde S_i = \tilde M_{i+1} \tilde L_{i+1} &  \\
& & & & \downarrow & \\
& & & & \frac{p_i}{q_{i+1}^2}S_{i+1} = \tilde L_{i+1} \tilde M_{i+1} & \rightarrow
\end{array}
\label{normal}
\end{equation}
where
$$
\tilde L_i=\frac{d}{dz}\frac{\sqrt{p_{i-1}}}{q_i},
\quad \tilde M_i=\frac{p_{i-1}^{3/2}}{q_{i}^3}\frac{d}{dz}\frac{q_{i}^2}{p_{i-1}},
\quad L_i=\frac{d}{dz}\frac{\sqrt{p_{i}}}{q_i},
\quad M_i=\frac{p_{i}^{3/2}}{q_{i}^3}\frac{d}{dz}\frac{q_{i}^2}{p_{i}}
$$
The horizontal arrows in (\ref{normal}) correspond to factorizations connected with
different choices of $p$ at fixed $q$ and vice versa,
while the vertical arrows correspond to permutation of factors. Due to the change of
(prefactors of) the second order operators
in the horizontal lines we do not have simple intertwining relations between
$S_0=d^2/dz^2$ and $S_i$:
$$
U^L_iS_0=S_iU^R_i
$$  
Note that now there are two different $U^L, U^R$ (left, right) intertwining operators instead of
one.
This makes the line of approach used for Adler-Moser polynomials unpromising for
$\Lambda=2$. 

\section*{V. Baker-Akhieser functions, charges in homogeneous field}

Again we ask the question: When the following indefinite integrals
\begin{equation}
\exp(-kz)\int\frac{p(k,z)}{q(k,z)^2}\exp(kz)dz,
\quad \exp(2kz)\int \frac {q(k,z)^4}{p(k,z)^2}\exp(-2kz)dz
\label{rationale}
\end{equation}
are rational? In (\ref{rationale}), $k$ is any complex number.

We call  the function
\begin{equation}
\Psi(k,z)=\frac{p(k,z)}{q(k,z)^2}\exp(kz)
\label{akhieser}
\end{equation}
rational Backer-Akhieser function for $\Lambda=2$ by analogy with $\Lambda=1$
(see below).

Similarly to the section III we can formulate the following

\begin{proposition}
Let polynomials $p(k,z), q(k,z)$ in $z$ do not have repeated or common roots.
The following three statements are equivalent:
\begin{itemize}
\item The indefinite integrals
$$
\exp(-kz)\int\Psi(k,z)dz , \quad \exp(2kz)\int\Psi(k,z)^{-2}dz
$$
are rational in $z$ functions, expressible without logarithms
\item
The energy function
\begin{equation}
\begin{array}{l}
E=k\sum_iQ_iz_i+\sum_{i<j}Q_iQ_j\ln|z_i-z_j|,\\
\\
Q_i=\left\{\begin{array}{l} 1, \quad i=1\dots n \\ -2,
\quad i=n+1\dots m+n \end {array} \right. , \quad n=\deg(p), m=\deg(q)
\end{array}
\label{equilibrium}
\end{equation}
has a critical point at
$$
z_i=\left\{\begin{array}{l} x_i(k), \quad i=1\dots n \\ y_{i-n}(k),
\quad i=n+1...m+n \end{array} \right.
$$
where
$x_i, y_i$ are roots of
$$
p=\prod_{i=1}^n\left(z-x_i(k)\right), \quad q=\prod_{i=1}^m\left(z-y_i(k)\right)
$$
\item
$p , q$ satisfy the following equation
\begin{equation}
0=\{p,q\}^{(k)}_2:=\{p,q\}_2+2k(p^\prime q-2q^\prime p)
\label{field}
\end{equation}
\end{itemize}
\label{eq}
\end{proposition}

Proof: Follows arguments similar to Lemmas \ref{lemmaeq}, \ref{lemmaint}.

One can easily see that equation (\ref{equilibrium}) describes equilibrium of two types of
charges with values 1, -2 in the homogeneous electric field.

Another observation consists in the fact that
\begin{equation}
p(k,z)=p(\zeta), \quad q(k,z)=q(\zeta), \quad \zeta=kz
\label{scale}
\end{equation}
Indeed, one can easily check this fact by scaling $z_i, i=1...n+m$ in (\ref{equilibrium}), or
by change the integration variable $z\to \zeta=kz$ in (\ref{rationale}).

Thus, we can set $k=1$ whenever $k\not=0$.

Equating highest symbols in (\ref{field}) we obtain the following simple

\begin{lemma}
Let $p, q$ satisfy conditions of proposition \ref{eq} then their degrees are related by
$$
deg(p)=n=2m=2\deg(q)
$$
\end{lemma}

In other words , the total charge has to be zero in order the system to be at rest in the
homogeneous field.

Although the case $\Lambda=2$ has some similarity with $\Lambda=1$, the same
approach to explicit representation
of $p_i, q_i$ , as shown in the section IV, does not apply any more.

Here we write down several first examples of $p_i(k,z), q_i(k,z)$
$$
q_0(k,z)=1, \quad p_0(k,z)=1,
$$
$$
q_1(k,z)=\zeta,  \quad p_1(k,z)=\zeta^2-3\zeta+3,
$$
\begin{equation}
\begin{array}{l}
q_2(k,z)=\zeta^3+t_2\zeta^2+\frac{t_2^2+6}{3}\zeta, \\
\\
p_2(k,z)=\zeta^6+\left (-9+2\,t_2\right )\zeta^5
+\left (40-15\,t_2+5/3\,{t_2}^{2}\right )\zeta^4 \\
\\
+\left (-96+52\,t_2-10\,{t_2}^{2}+2/3\,{t_2}^{3}
\right )\zeta^3+\left (112-90\,t_2+{\frac {76}{3}}\,{t_2}^{2}-3\,{t_2}^{3}
+1/9\,{t_2}^{4}\right )\zeta^2 \\
\\
+
\left (-48+66\,t_2-28\,{t_2}^{2}+5\,{t_2}^{3}-1/3
\,{t_2}^{4}\right )\zeta-18\,t_2+10\,{t_2}
^{2}-3\,{t_2}^{3}+1/3\,{t_2}^{4}+48
\end{array}
\label{t2}
\end{equation}
where $t_2$ is an arbitrary constant.

It is easy to show that solution (\ref{t2}) is generic for $m=3$ , in the sense that it is not contained
in a wider class. In other words , coefficients of (properly normalized) $q_2$ cannot depend on more than
one free parameter. Indeed , we can always take $q_2$ to be monic and without constant term (by fixing one
root at $z=0$). If $q_2$ depended on more than one parameter, then it would be $q_2=z^3+t_2z^2+Tz$, where
$t_2$ and $T$ are arbitrary constants. In such a case (\ref{field}) become a linear equations for the coefficients
of $p_2$.
It is verified by elementary linear algebra that this system is incompatible. Thus $q_2$ can depend only on one free parameter (say $t_2$).

It is difficult to get examples of polynomials of higher degrees, even with the help
of the computer.
It is also nontrivial problem to define degrees of $q(k,z)$.
If one assumes that by analogy with $\Lambda=1$ case the order of $q_i(k,z)$
is a quadratic function of $i$ (which certainly holds for
$k=0$, see proposition \ref{nofield}),
then it follows from the above examples that
$\deg(q_i(k,z))=i(i+1)/2$.

\section*{VI. Scaling, reparametrization and bispectrality}

A statement similar to proposition \ref{eq} holds for the case $\Lambda=1$ with the Baker-Akhieser function
$\psi=(p/q)\exp(kz)$ instead of (\ref{akhieser}) and symmetric integration conditions
$$
\exp(-kz)\int \psi dz, \quad \exp(kz)\int \psi^{-1}dz
$$
The electrostatic analogy describes charges 1, -1 in the homogeneous field and equation
(\ref{field}) is replaced by
$$
0=\{p,q\}_1+2k(p^\prime q-q^\prime p)
$$
This equation can be rewritten in the Schr\"odinger form
\begin{equation}
L\psi(k,z)=\left(\frac{d^2}{dz^2}+V\right)\psi(k,z)=k^2\psi(k,z), \quad V=2\frac{d^2}{dz^2}\log q
\label{units}
\end{equation}
It follows from (\ref{up}), (\ref{shift}) (for more details see \cite{KDV}) that
the solution of (\ref{units}) is obtained by action of the intertwining operator (\ref{shift}) to the
eigenfunction $\exp(kz)$ of the free Schrodinger operator $d^2/dz^2$. This solution is the Backer-Akhieser function:
$$
\psi=W[\psi_1,\dots,\psi_n,\exp(kz)]/W[\psi_1,\dots,\psi_n]
$$
Therefore, according to (\ref{polynomials})
$$
p(k,z)=\exp(-kz)W[\psi_1,\dots,\psi_n,\exp(kz)], \quad
q(k,z)=W[\psi_1,\dots,\psi_n]={\rm const}\theta_n(z)
$$
On the other hand , by arguments leading to (\ref{scale}) , we see that
\begin{equation}
\left(\frac{d^2}{d\zeta^2}+V\right)\psi(\zeta)=\psi(\zeta), \quad V=2\frac{d^2}{d\zeta^2}\log q
\label{zeta}
\end{equation}
with
$$
\psi(\zeta)=\frac{p(\zeta)}{q(\zeta)}\exp(\zeta), \quad p(\zeta)=\exp(-\zeta)W[\psi_1(\zeta),\dots,\psi_n(\zeta),\exp(\zeta)], \quad
q(\zeta)={\rm const}\theta_n(\zeta)
$$
We observe that potential $V=V(z)$ in (\ref{units}) is independent of $k$, while potential
$V=V(\zeta)$, in (\ref{zeta}) , depends on $k$. This is due to the fact that
rescaling of variable $z\to \zeta=kz$ can be absorbed by appropriate changes of the
parameters (see e.g. (\ref{kdv}))
$$
\theta_n(z; t_1, t_2, ...,t_n)=k^{-n(n+1)/2}\theta_n(kz; kt_1, k^3t_2, ..., k^{2n-1}t_n )
$$
since the Adler-Moser polynomials can be seen as homogeneous polynomials of variables
$z$ and $t_i, i>0$ with weights 1 and $2i-1$ correspondingly.

Thus the problem of equilibrium of charges $1$ and $-1$ can be reduced to the spectral problem
for the $k$-independent operator $L$ (\ref{units}).

Similarly, one can ask if $\psi(k,z)$ is simultaneously a solution of a spectral problem
for $z$ independent differential with respect to $k$ operator $A$
$$
A\psi(k,z)=\Theta(z)\psi(k,z), \quad \partial A/\partial z = 0
$$
where $\Theta(z)$ is a function of $z$.

It turns out \cite{BI} that such operators exist and belong to the so called odd bispectral family.

In general, the problem of finding functions satisfying simultaneously a differential equation in $z$ with $k$-dependent eigenvalues and a differential equation in $k$ with $z$-dependent eigenvalues is called the bispectral problem \cite{BI}.

Now returning to the main subject of this work , one might expect that the $\Lambda=2$
case can belong to some other bispectral (e.g. even \cite{BI}) family of the differential operators.
Unfortunately , the $\Lambda=2$ case is not , in general , related to the bispectral problem for a second order differential operator. Indeed , equation (\ref{field}) can be rewritten in the following two forms
\begin{equation}
\begin{array}{l}
\left(\frac{d^2}{d\zeta^2}+u(\zeta)\right)\Psi(\zeta)=\Psi(\zeta), \quad u(\zeta)=6\frac{d^2}{d\zeta^2}\log q(\zeta)\\
\\
\left(\frac{d^2}{d\zeta^2}+v(\zeta)\right)\Psi(\zeta)^{-1/2}=\Psi(\zeta)^{-1/2}, \quad v(\zeta)=\frac{3}{4}\frac{d^2}{d\zeta^2}\log p(\zeta)
\end{array}, \quad \Psi(\zeta)=\frac{p(\zeta)}{q(\zeta)^2}\exp(\zeta)
\label{uv2t}
\end{equation}
We use (\ref{t2}) as a counterexample to the bispectrality. It is seen that there is
no change of parameter $t_2\to f(t_1,t_2)$ combined with the translation 
$z\to z+t_1$ in (\ref{t2}) , such that $p_2$ or $q_2$ become homogeneous (with some weights) in $z$ and $t_2$. Thus, in difference from the $\Lambda=1$ case, rescaling of variable $z$ cannot be absorbed by a change of parameter $t_2$. As a consequence , potentials $u, v$ in (\ref{uv2t}) cannot be $z$ or $k$ independent. Therefore , it is impossible , in general , to set a bispectral problem with the second order differential operator in $z$ (or $k$) when $\Lambda=2$. 

\section*{VII. Integrable dynamics of charges}

The charge configurations studied above can be considered as fixed points of some
dynamical systems \cite{L}.

\begin{lemma}

Let two polynomials $p(z,t)=\prod_{i=1}^n(z-z_i(t))$ and
$q(z,t)=\prod_{i=n+1}^{n+m}(z-z_i(t))$ satisfy the following bilinear equation
\begin{equation}
q\frac{dp}{dt}-\Lambda p\frac{dq}{dt}=\{p,q\}_\Lambda
\label{bilinear}
\end{equation}
Then the roots $z_i(t), i=1\dots n+m$ satisfy the system of ordinary differential
equations
\begin{equation}
\frac{dz_i}{dt}=\sum_{j\not=i=1}^{n+m}\frac{Q_j}{z_i-z_j} = \frac{1}{Q_i} \frac{\partial E}{\partial z_i}
\label{vortices}
\end{equation}
where
$$
Q_i=\left\{\begin{array}{l} 1, \qquad i=1 \dots n \\ -\Lambda, \qquad i=n+1
\dots m+n \end{array}\right.
$$
\end{lemma}

It follows from the above lemma that the critical points of the energy (\ref{energy}) are fixed points of
(\ref{vortices}). Although, as seen from the previous considerations,
existence of such critical points depends on values $m, n$ and $\Lambda$, equation
(\ref{bilinear}) has polynomial solutions for any integer $m,n\ge 0$ and real $\Lambda$.  

Although (\ref{vortices}) (or equivalently (\ref{bilinear}) ) is not a Hamiltonian system,
it can be embedded in the Hamiltonian flow:

\begin{proposition}
(\ref{vortices}) is a trajectory of a system with the Hamiltonian
\begin{equation}
H=\sum_{i=1}^{n+m}\left(\frac{dz_i}{dt}\right)^2- \sum^{n+m}_{i<j=1}  \frac{Q_iQ_j(Q_i+Q_j)}{(z_i-z_j)^2}
\label{new}
\end{equation}
In other words, the Hamiltonian equations of motion
$$
\frac{d^2z_i}{dt^2}=\sum_{j\not=i=1}^{n+m}\frac{Q_iQ_j(Q_i+Q_j)}{(z_i-z_j)^3}
$$
are corollaries of (\ref{vortices})
\end{proposition}

Proof: This proposition is a corollary of more general result proved in \cite{L} .

This is a nontrivial fact that the Hamiltonian system of the point masses interacting through two body forces is a corollary of equations of motion of a lower order for pairwise interacting points. In general , such a property is equivalent to compatibility of a highly overdetermined system of equations. Perhaps , elliptic generalizations of (\ref{vortices}) are the only systems satisfying such conditions \cite{L}.

It is easy to note that, when $\Lambda=1$ (\ref{new}) is a sum of two
independent Hamiltonians
$$
H=H_1+H_2, \quad 
H_1=\sum_{i=1}^n\left(\frac{dz_i}{dt}\right)^2-\sum_{i<j=1}^n\frac{2}{(z_i-z_j)^2}, 
\quad H_2=\sum_{i=n+1}^{n+m}\left(\frac{dz_i}{dt}\right)^2-
\sum_{i<j=n+1}^{n+m}\frac{2}{(z_i-z_j)^2}
$$
Each of them belongs to the completely integrable
(in the Liouville sense) Calogero-Moser system \cite{M}. As a corollary
(\ref{vortices}) is integrable when $\Lambda=1$  .

Equation (\ref{new}) is also the Calogero Moser Hamiltonian if $\Lambda=-1$.

In general, (\ref{new}) is not reduced to known integrable Hamiltonians.

It is conjectured in \cite{L} that (\ref{vortices}) is integrable for any real
$\Lambda$ in
the sense that there exist $2(n+m)-1$ functionally independent real constants of motion
which are rational functions of $z_i$ and $z^*_i$ (asterisk denotes the complex
conjugation and $i=1\dots n+m$).

It can be seen from the above the case $\Lambda=1$ has a particular significance
both in
the problem of the rational integrals (\ref{rational}) and from the point of view of the
integrable Hamiltonian systems. It is interesting to establish the role of the second
case
$\Lambda=2$ in the theory of the dynamical system (\ref{vortices}).

\section*{VIII Conclusion}

The principal question of this paper to be addressed is about existence and explicit representation
of the Baker-Akhieser function (or equivalently $p(k,z),q(k,z)$) for $\Lambda=2$.
It is worth to mention a useful generalization (interesting by itself)
which might help to find the answer:
By analogy with the $\Lambda=1$ case, one may consider
equilibrium of charges on the cylinder. Remind that (e.g. see \cite{B},\cite{L}),
equilibrium configurations for $\Lambda=1$ are zeros of
trigonometric polynomials obtained by finite
numbers of Darboux transformations from the free Schr\"odinger operator on circle.
In such setting, both the Adler-Moser
polynomials and $\Lambda=1$ rational Baker-Akhieser function can be obtained choosing
different 
parameters in the limit when the radius of the cylinder goes to infinity.

It might be interesting to examine a similar procedure getting an analog of proposition \ref{nofield} on the cylinder.

Another interesting question , as was mentioned before , is to understand the role of the case $\Lambda=2$
in the theory of dynamical systems (\ref{vortices}) and integrability of related Hamiltonians.

Finally , we would like to mention relation with the Reduced Model of Superconductivity (Richardson, \cite{BCS}) , the Gaudin Magnet Model \cite{Magnet} from the one hand and the $\Lambda=2$ equilibrium configurations from the other hand. These models are sets of commuting quantum Hamiltonians acting on a finite-dimensional Hilbert space. Equations for the Hamiltonian eigenvalues are equilibrium conditions for a set of charges of value $-2$
located in the complex plane, subject to mutual repulsion , and attraction of charges of value $1$ located at positions defined by the parameters of the Hamiltonian. The $\Lambda=2$ case considered in the present work corresponds to the Richardson (Gaudin) models with special restrictions imposed on the parameters of Hamiltonian.It is interesting to examine relations between such models and establish their special properties connected with the $\Lambda=2$ case.

\section*{Acknowledgments}

The author is grateful to H.Aref, Y.Berest, B.Dubrovin, T.Grava, O.Yermolayeva for useful information , help and remarks.

\end{document}